\documentclass[iop]{emulateapj}
\usepackage{apjfonts}
\usepackage{graphicx}

\newcommand{\be}{\begin{equation}}
\newcommand{\ee}{\end{equation}}
\newcommand{\Msun}{M_{\odot}}

\def\Lsun{\,L_\Sol}
\def\Msun{\,M_\Sol}
\newcommand{\Mbh}{M_{\rm BH}}
\newcommand{\Mgal}{M_{\rm gal}}

\shorttitle{Quasar Demographics}
\shortauthors{CONROY \& WHITE}

\begin{document}
\journalinfo{The Astrophysical Journal}

\title{A Simple Model for Quasar Demographics}

\author{Charlie Conroy$^1$ \& Martin White$^{2,3}$}

\affil{$^1$Department of Astronomy \& Astrophysics,
  University of California, Santa Cruz, CA, 95060, USA \\ 
$^2$Departments of Physics and Astronomy,
  University of California, Berkeley, CA 94720, USA \\
$^3$Physics Division, Lawrence Berkeley National Laboratory,
  Berkeley, CA 94720, USA}

\slugcomment{Submitted to ApJ}

\begin{abstract}

  We present a simple model for the relationship between quasars,
  galaxies, and dark matter halos from $0.5<z<6$.  In the model, black
  hole (BH) mass is linearly related to galaxy mass, and galaxies are
  connected to dark matter halos via empirically constrained
  relations.  A simple ``scattered'' light bulb model for quasars is
  adopted, wherein BHs shine at a fixed fraction of the Eddington
  luminosity during accretion episodes, and Eddington ratios are drawn
  from a lognormal distribution that is redshift-independent.  This
  model has two free, physically meaningful parameters at each
  redshift: the normalization of the $\Mbh-\Mgal$ relation and the
  quasar duty cycle; these parameters are fit to the observed quasar
  luminosity function (LF) over the interval $0.5<z<6$.  This simple
  model provides an excellent fit to the LF at all epochs, and also
  successfully predicts the observed projected two-point correlation
  of quasars from $0.5<z<2.5$.  It is significant that a {\it single}
  quasar duty cycle at each redshift is capable of reproducing the
  extant observations.  The data are therefore consistent with a
  scenario wherein quasars are equally likely to exist in galaxies,
  and therefore dark matter halos, over a wide range in masses.  The
  knee in the quasar LF is a reflection of the knee in the stellar
  mass-halo mass relation.  Future constraints on the quasar LF and
  quasar clustering at high redshift will provide strong constraints
  on the model.  In the model, the autocorrelation function of quasars
  becomes a strong function of luminosity only at the very highest
  luminosities, and will be difficult to observe because such quasars
  are so rare.  Cross-correlation techniques may provide useful
  constraints on the bias of such rare objects.  The simplicity of the
  model allows for rapid generation of quasar mock catalogs from
  N-body simulations that match the observed luminosity function and
  clustering to high redshift.

\end{abstract}



\section{Introduction}
\label{sec:introduction}

Quasars are among the most luminous astrophysical objects, and are
believed to be powered by accretion onto supermassive black holes
\citep[e.g.][]{Sal64,Lyn69}.  They have become a key element in our
current paradigm of galaxy evolution \citep[e.g.,][]{Spr05, Croton06,
  Hop08}, and essentially all spheroidal systems at present harbor
massive black holes \citep{KorRic95}, the masses of which are
correlated with many properties of their host systems.  Despite their
importance, and intense theoretical activity, a full theory of the
coevolution of galaxies and quasar eludes us.

The current paradigm assumes that every galaxy initially forms in a
gas-rich, rotationally-supported system.  Once the dark matter halo
grows to a critical scale some event, most likely a major merger
\citep{Car90, HaiLoe98, CatHaeRee99, KauHae00, Spr05, Hop06, Hop08} or
instability in a cold-stream fed disk \citep{DiM12}, triggers a period
of rapid, obscured star formation, the generation of a stellar bulge
and a growing black hole (BH).  Eventually the accreting BH becomes
visible as a quasar, and soon after the star formation is quenched on
a short timescale, perhaps via radiative or mechanical feedback from
the BH \citep[e.g.][]{Sil98, Kin03, WyiLoe03, Sha09, Nat12, AleHic12}.
Understanding the details of this picture remains an active area of
research.

Phenomenological models for quasar demographics often adopt power-law
relations between quasars, galaxies, and dark matter halos
\citep[e.g.,][]{EfsRee88, Car90, WyiLoe02, WyiLoe03, HaiCioOst04,
  Mar06, Lid06, Croton09, She09, BooSch10}.  In these models, the duty
cycle of quasars is tuned to match the observations, and a generic
conclusion is that the duty cycle is a strong function of halo mass or
quasar luminosity, peaking at a halo mass of $10^{12-13}\Msun$.
However, these previous models do not incorporate constraints provided
by the galaxy stellar mass function over the interval $0<z<6$.  And
yet, a variety of lines of evidence suggest that the relation between
halos and galaxies is highly non-linear, with a characteristic peak in
galaxy formation efficiency at a halo mass of $\sim10^{12}\Msun$
\citep{Van03, Val04, Man06, Con09, Mos10, TruKlyPri11, BehWecCon12}.
The aim of this paper is to incorporate empirically constrained
relations between galaxies and halos into a simple model for quasar
demographics.  We will demonstrate that a model constructed to match
the observed galaxy stellar mass function implies a quasar duty cycle
that is independent of galaxy and halo mass at each redshift.  This
has important implications for physical models aimed at understanding
the triggering of quasars and their connection to the evolution of
galaxies.

The outline of the paper is as follows.  In \S\ref{sec:model} we
describe the model, in \S\ref{sec:data} the model is compared to data,
and a discussion is presented in \S\ref{sec:discussion}.  We conclude
in \S\ref{sec:conclusions}.  Where necessary we adopt a $\Lambda$CDM
cosmological model with $\Omega_m=0.28$, $\Omega_\Lambda=0.72$ and
$\sigma_8=0.8$.  Unless the $h$ dependence is explicitly specified or
parametrized, we assume $h=0.7$.  Dark matter halo masses are quoted
as $M_{\rm vir}$ \citep{Bryan98}.  Luminosities are quoted in Watts
and magnitudes in the AB system, and stellar masses assume a
\citet{Cha03} stellar initial mass function.


\section{The model}
\label{sec:model}

Our goal is to construct a simple model that relates galaxies,
quasars, and dark matter halos over the redshift interval $0<z<6$.  A
small number of free parameters will characterize the model, and these
parameters will be constrained against observations.

The most constraining observation will be the quasar luminosity
function, and to predict that in our model we could begin with the
observed stellar mass function.  However it will be useful later to
have information on how quasars occupy dark matter halos, and for this
reason we begin by specifying a dark matter halo mass function and its
evolution to $z=6$.  We adopt the fitting functions of
\citet{Tin08,Tin10} for the halo mass function and large-scale bias,
which represent the latest fits to these parameters from cosmological
$N-$body simulations\footnote{The fits are only calibrated to $z=2$,
  but we checked the mass function fit agrees with our $N$-body
  simulation to better than a factor of 2 up to $z=6$.}.  Note that
here and throughout we consider only parent halos; satellite halos,
also known as subhalos, are not included in the present study.  This
is a reasonable approximation at high redshift, as quasars inhabit
highly biased halos on the steeply falling tail of the mass function
and any satellite galaxies of the same mass would live in even more
massive halos which are exponentially rare.  This assumption will
break down at lower luminosities, where the satellite fraction can be
expected to rise.  This assumption will also fail to account for the
small-scale clustering of quasars, in particular the clustering within
the halo scale of $\lesssim1$ Mpc.  When we compare to clustering
measurements in $\S$\ref{sec:clust} we will therefore restrict our
comparison to $R>1\,$Mpc, which is where most of the data lie.
Extending the model to satellites is in principle straightforward, but
requires an assumption about the joint occupation of quasars in
central and satellite galaxies of the same halo.

We adopt empirically constrained relations between galaxy stellar mass
and dark matter halo mass over the interval $0<z<6$ from
\citet{BehWecCon12}.  Briefly, these relations were constrained by
populating dark matter halo merger trees with galaxies via
redshift-dependent $M_h-\Mgal$ relations.  Model galaxy stellar mass
functions were then computed by taking into account observational
uncertainties in the stellar mass estimates and galaxy star formation
rates were computed by following the growth of galaxies through the
merger trees.  The model stellar mass functions and star formation
rate functions were compared to a comprehensive compilation of
observations.  The underlying $M_h-\Mgal$ relations were varied until
a good match to the data was achieved.  The resulting relations agree
with results obtained from other techniques, including abundance
matching, halo occupation models, satellite kinematics, and
gravitational lensing \citep[see][]{BehWecCon12}.  We also adopt an
amount of scatter between galaxy mass and halo mass as a function of
redshift implied by the model of \citet{BehWecCon12}.  This scatter
increases from $\approx0.2$ dex at $z=0.5$ to $\approx0.5$ dex at
$z=6$, although some of this `scatter' reflects observational
uncertainty.

Galaxies are assigned BHs via the following equation:
\begin{equation}
 \frac{\Mbh}{10^{10}M_\odot} =
 10^\alpha\,(1+z)^2\,
 \left(\frac{\Mgal}{10^{10}M_\odot}\right)^\beta \,,
\label{eqn:mbh_mgal}
\end{equation}
where $\Mgal$ and $\Mbh$ are the stellar mass of the galaxy and mass
of the BH, respectively.  The available data at $z\sim0$ is consistent
with a linear relation between $\Mgal$ and $\Mbh$, (i.e.~$\beta=1$)
which is what we adopt herein, with a normalization constant of
$\alpha\approx-3.1$ \citep{HarRix04}.  The scaling with redshift is
motivated by observations \citep{McL06,Tar12}, but since we fit for
$\alpha$ at each redshift, any deviation from $(1+z)^2$ will be
absorbed in the redshift-dependence of the parameter $\alpha$.  In our
fiducial model we adopt a scatter in this relation of 0.3 dex,
independent of mass,  consistent with the observed scatter in the local
$\Mbh-\sigma$ relation \citep{Tre02}.  

We have chosen to relate $\Mbh$ to the total stellar mass of the
galaxy, rather than specifically to the bulge component.  Obviously
for bulge-dominated galaxies the distinction is irrelevant, but the
differences can grow as we include galaxies with a large disk
component.  Assuming that bulge properties are the dominant factor in
determining $\Mbh$, a more refined model would include the evolution
and mass-dependence of the bulge-to-total ratio.  However for now we
neglect this distinction.  We do find that our results are relatively
robust to modest changes in the slope of the $\Mbh-\Mgal$ relation
(see \S\ref{sec:data}) --- and any overall normalization change can be
absorbed into our parameter $\alpha$ --- so there are reasons to
believe a more complex\footnote{Such a model might couple $\Mbh$ to
  $\Mgal\simeq M_{\rm bulge}$ at high-$z$ but allow low-$z$ galaxies
  to (re)grow disks leading to evolution in $\Mbh-\Mgal$ but not
  $\Mbh-M_{\rm bulge}$, see
  e.g. \citet{Jah09,Cis11,KorBen11,KorBenCor11}.}  model would achieve
a similar level of success in fitting the observations.

In addition to the strong observed correlation between $\Mgal$
and $\Mbh$, there are well-known correlations between $\Mbh$ and other
parameters of the galaxy including the velocity dispersion, $\sigma$,
and galaxy size, $R_e$.  In fact, \citet{Hop07b} argued for the
existence of a BH fundamental plane (relating $\Mbh$, $\sigma$, and
$R_e$) that has smaller scatter than any other relationship between
$\Mbh$ and a single galaxy property.  Another option would therefore
have been to connect BHs to galaxies via $\sigma$, as for example done
by \citet{Croton09}, or via the BH fundamental plane.  We choose to
use $\Mgal$ herein because this quantity is readily available
for galaxies to $z=6$, and because the redshift-dependent connection
between galaxies and halos is presently available for galaxy stellar
masses, but not for galaxy velocity dispersions.

The BH mass is converted to a bolometric quasar luminosity through the
Eddington ratio, $L/L_{\rm Edd}\equiv\eta$,
\begin{equation}
  L_Q = 3.3\times10^4\ \eta\  \frac{\Mbh}{\Msun}\, \Lsun.
\label{eqn:Lq_Mbh}
\end{equation}
In our fiducial model $\eta$ is independent of redshift.  We draw
$\eta$ from a lognormal distribution with mean of $\eta=0.1$ and a
dispersion of $0.3\,$dex, in agreement with observations \citep{Kol06,
  She08}.  In our model the value of the Eddington ratio is degenerate
with the normalization of the $\Mbh-\Mgal$ relation and any intrinsic
width in the Eddington ratio distribution is degenerate with scatter
in the $\Mbh-\Mgal$ relation.  In order to explore this degeneracy we
consider a second model where $\eta$ is 0.1 at low redshift, increases
linearly between $0.5<z<3.5$ to a value of 1.0, and at higher
redshifts $\eta=1.0$ \citep[see e.g.,][]{Wil10b, She12}.  These two
models will serve to indicate a reasonable range in possible evolution
in the Eddington ratio.

\begin{figure}[!t]
\begin{center}
\resizebox{3.5in}{!}{\includegraphics{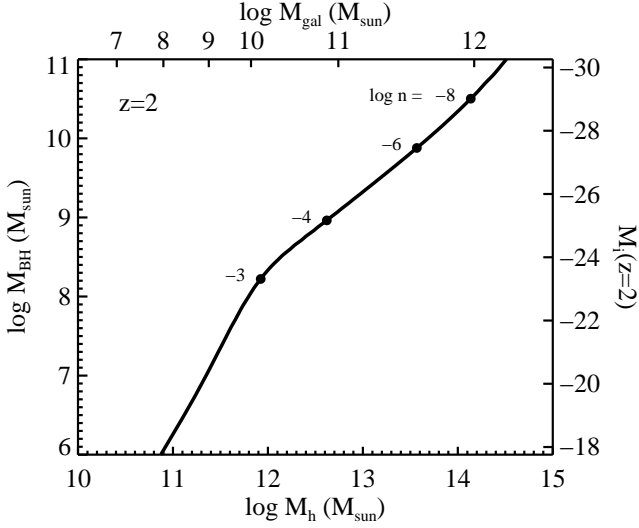}}
\end{center}
\caption{Summary of the model relations at $z=2$.  The quasar LF
  determines the abundance (see the points on the curve, which label
  space densities in units of log Mpc$^{-3}$) of quasars at a given
  luminosity (right vertical axis) or BH mass (left vertical axis).
  For an assumed lifetime, $t_Q$, this maps to an abundance of
  galaxies and the stellar mass function provides the appropriate
  galaxy stellar mass (upper horizontal axis).  The empirically
  constrained $\Mgal-M_h$ relations from \citet{BehWecCon12}
  allow us to map this into a halo mass (lower horizontal axis).  The
  curve shown is at $z=2$, though the general behavior is similar at
  other redshifts with a steep low-mass slope and a shallower high
  mass slope (see Figure \ref{fig:mbh_mhalo}).  Note the lower
  horizontal axis determines the clustering amplitude at fixed
  redshift while the left vertical axis determines the quasar
  luminosity.}
\label{fig:QAM}
\end{figure}

In order to compare to observations, we must translate $L_Q$ into
magnitudes in a given filter.  We adopt the relation between
bolometric luminosity and $i$-band magnitude (k-corrected to $z=2$)
using the relation from \citet{Shen++09}:
\begin{eqnarray}
M_{i}(z=2) &=& \hphantom{-}72.5 - 2.5\,\log L_Q \label{eq:lboldef} \\
&=& -5.26 - 2.5\,\log\left(\eta \Mbh\right) \\
&=& -30.3 - 2.5\,\left(\log\eta + \alpha\right) - 5\log(1+z) \nonumber \\
& & - 2.5\beta\,\log\left(\Mgal/10^{10}M_\odot\right) \label{eq:lboldef2}
 \,\,,
\end{eqnarray}
where $L_Q$ is in Watts and $\Mbh$ is in solar masses.  The last two
relations follow directly from Equations \ref{eqn:mbh_mgal} and
\ref{eqn:Lq_Mbh}; we include them here to make explicit the connection
between $\Mgal$ and observed quasar magnitude, and also to emphasize
the fact that $\eta$ and $\alpha$ are perfectly degenerate in our
model.  There is scatter in $L_Q$ at fixed $\Mgal$ which arises from a
combination of scatter in $\Mbh-\Mgal$ and $L_Q-\Mbh$.  In our model
we adopt a scatter of 0.3 dex between each of these relations,
resulting in a total scatter between $\Mgal$ and $L_Q$ of 0.42 dex.

There are two free parameters in this model at each redshift: the
normalization of the $\Mbh-\Mgal$ relation, specified by $\alpha$, and
the quasar duty cycle, $f_{\rm on}$.  These two parameters are fit to
the observed quasar LF via $\chi^2$ minimization.  An important, and
novel feature of this model is that we adopt a constant duty cycle,
independent of luminosity, $\Mbh$ or $M_h$.  Sometimes the duty cycle
is recast into a ``lifetime'' using the Hubble time: $t_Q\equiv f_{\rm
  on}t_H$.  As we will demonstrate in the following section, both of
these parameters are highly constrained by the observed quasar LF.

The resulting relations between galaxies, halos, and quasars are
illustrated in Figure \ref{fig:QAM}.  These relations represent the
best-fit model constrained by the quasar LF at $z=2$ (see
$\S$\ref{sec:lf}).  The quasar LF allows us to relate luminosity to
number density.  For an assumed duty cycle we then have the abundance
of BHs of that mass.  Similarly the stellar mass function maps galaxy
mass to abundance.  Thus at fixed duty cycle we obtain a tight
constraint on $\Mbh-\Mgal$.  As the stellar mass function and quasar
LF contain significant curvature only one combination of normalization
and duty-cycle provides a good fit to the data for a range of
luminosities (unless we allow significant variation in the lifetime as
a function of luminosity).

\begin{figure}[!t]
\begin{center}
\resizebox{3.5in}{!}{\includegraphics{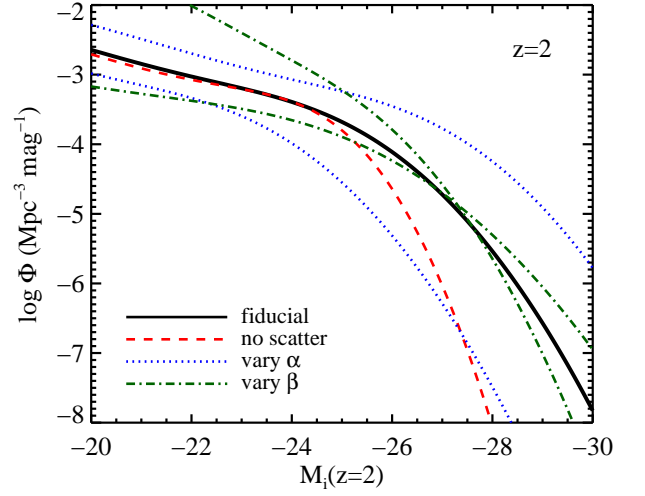}}
\end{center}
\caption{Variation in the predicted luminosity function of quasars at
  $z=2$ as a function of the parameters in our model.  The dashed
  (red) line shows how the inclusion of scatter in the $\Mbh-M_{\rm
    gal}$ relation is important at the high mass end, with models
  including more scatter predicting more luminous quasars.  Variations
  due to changes in the normalization of the $\Mbh-\Mgal$ relation
  ($-3.4<\alpha<-2.8$; Equation \ref{eqn:mbh_mgal}) are shown by the
  dotted (blue) lines, and we see this parameter changes both the
  normalization and shape of the LF since the galaxy stellar mass
  function has a particular shape.  Finally the dot-dashed (green)
  line shows variation in the logarithmic slope of the $\Mbh-M_{\rm
    gal}$ relation ($0.5<\beta<1.5$; Equation \ref{eqn:mbh_mgal}).}
\label{fig:lf_params}
\end{figure}

Figure \ref{fig:lf_params} shows how the predicted quasar luminosity
function at $z=2$ depends upon several parameters in the model.  The
amount of scatter in the $L_Q-\Mgal$ relation is important for the
shape at high luminosity, and indeed the abundance of luminous quasars
provides a lower limit on the scatter for any model which places
quasars in halos on the exponentially falling part of the mass
function.  We see that a model with no scatter in the $L_Q-\Mgal$
relation predicts drastically fewer bright quasars and a steeper
bright-end slope than a model including scatter \citep[see also][for
related discussion]{WhiMarCoh08, ShaWeiShe10, DeGraf11, TraSte12}.
Variations in the BH mass at fixed galaxy mass ($\alpha$) change both
the normalization and shape of the luminosity function while variation
in the slope of the relation ($\beta$) has a large effect on the shape
of the LF both at low and high luminosity.

\begin{figure*}[!t]
\begin{center}
\resizebox{6.5in}{!}{\includegraphics{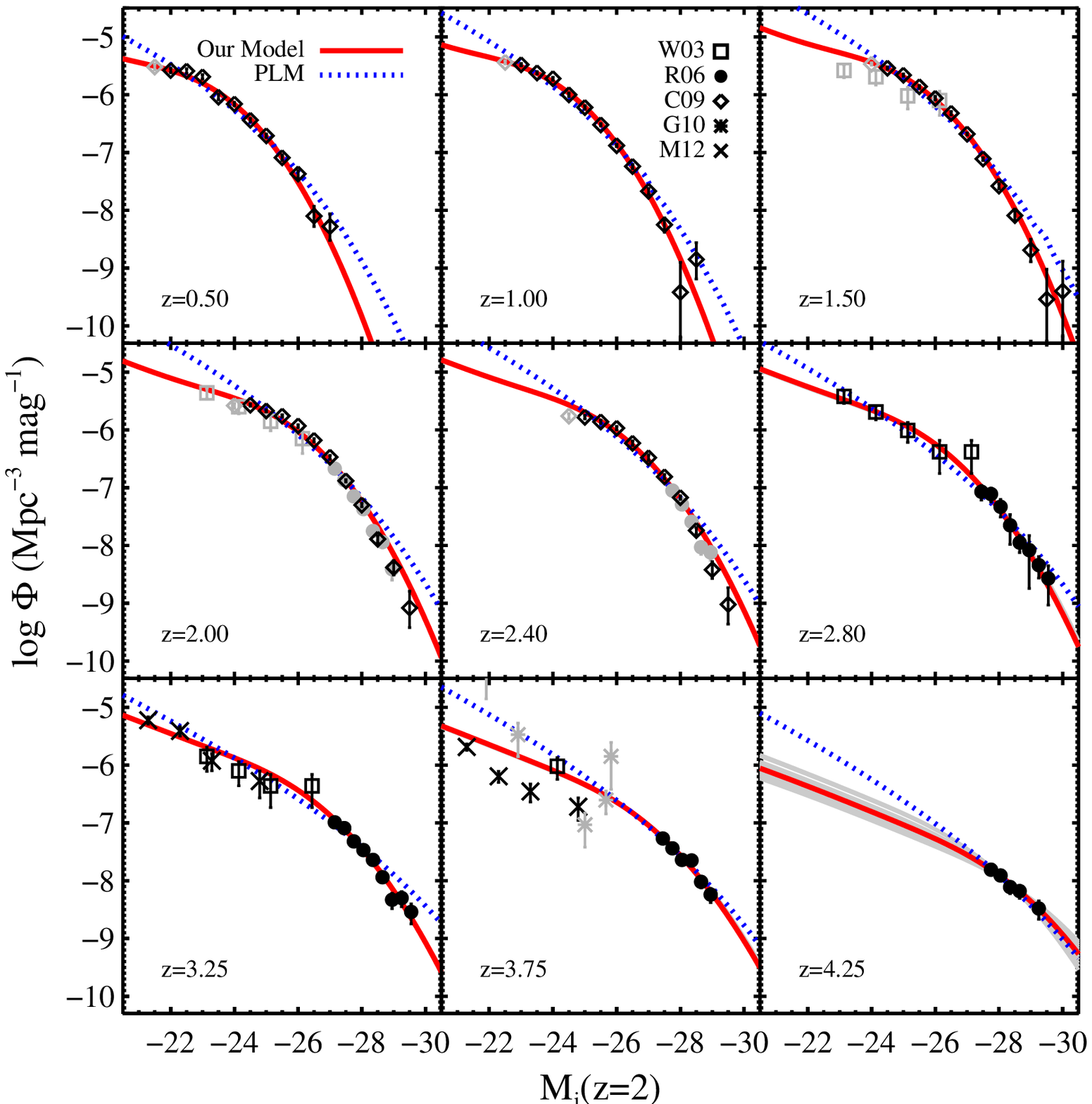}}
\end{center}
\caption{The quasar luminosity function predicted by our model at
  different redshifts, as compared to the observations and a simple
  model in which quasar luminosity is tied to halo, not galaxy, mass
  (denoted PLM for power-law model).  The data are from
  \citet[][COMBO-17; open squares]{Wol03}, \citet[][SDSS; solid
  circles]{Ric06}, \citet[][2SLAQ+SDSS; open diamonds]{Cro09},
  \citet[][NDWFS+DLS; stars]{Gli10}, and \citet[][COSMOS;
  crosses]{Mas12}.  The lifetime, $t_Q$, and the $\Mbh-\Mgal$
  normalization, $\alpha$, are fit in each panel and the grey region
  illustrates the $1\,\sigma$ uncertainty in the model prediction.
  Only black symbols are included in the fits; the grey symbols
  generally represent data of lower quality and are included for
  comparison purposes only.}
\label{fig:lf}
\end{figure*}

\begin{figure}[!t]
\begin{center}
\resizebox{3.5in}{!}{\includegraphics{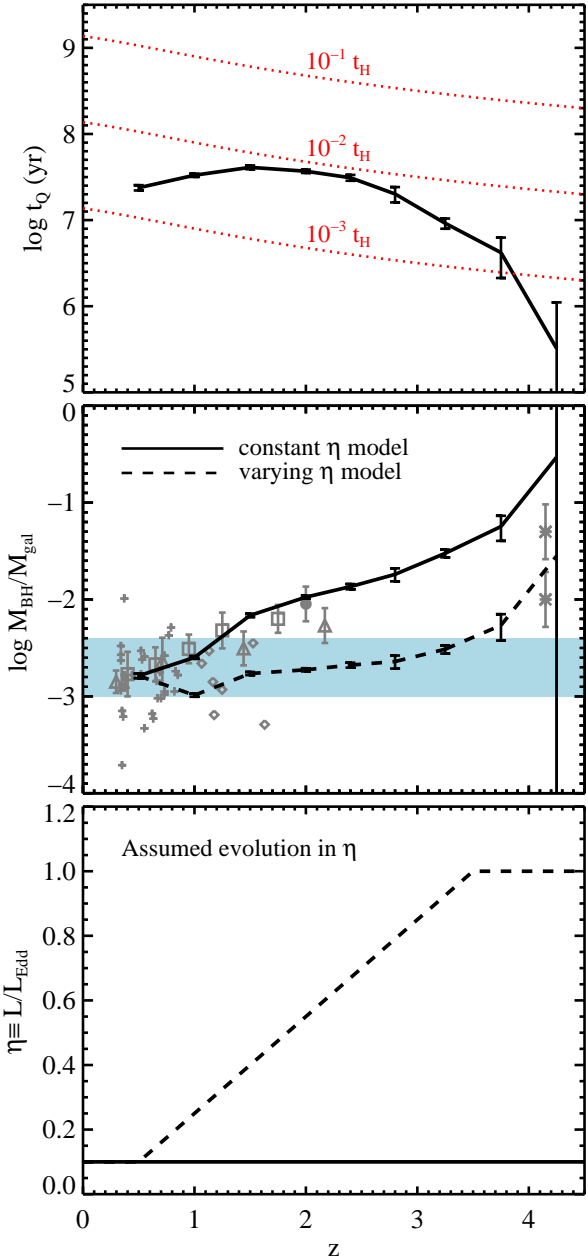}}
\end{center}
\caption{Upper Panel: The duty-cycle, or quasar lifetime, as a
  function of redshift.  We define $t_Q=f_{\rm on}t_H$ where $t_H$ is
  the Hubble time at redshift $z$ and $f_{\rm on}$ is the probability
  that a BH is a luminous quasar (which is independent of luminosity
  in our model).  Also shown are lines of constant $f_{\rm
    on}=10^{-1}$, $10^{-2}$ and $10^{-3}$.  Middle Panel: Evolution of
  the normalization of the $\Mbh-\Mgal$ relation in our model (for two
  choices of evolution in $\eta$; solid and dashed lines) compared to
  results from the literature.  The solid band is the normalization at
  $z=0$ \citep{HarRix04}.  Plus symbols and diamonds are individual
  measurements from \citet{Cis11} and \citet{Jah09}, respectively.
  Triangles are binned estimates from \citet{Dec10}, squares are
  binned estimates from \citet{McL06}, the solid circle is a binned
  measurement from \citet{Pen06}, and stars are the average of two
  quasars from \citet{Tar12} for two choices for estimating galaxy
  masses. Lower Panel: Assumed evolution in the Eddington ratio,
  $\eta$, for the two models shown in the middle panel.}
\label{fig:dutycycle}
\end{figure}


\vspace{2cm}

\section{Comparison with observational data}
\label{sec:data}

\subsection{The Quasar Luminosity Function}
\label{sec:lf}

Figure \ref{fig:lf} shows the predictions of our model compared to a
compilation of observational data from \citet[][COMBO-17; open
squares]{Wol03}, \citet[][SDSS; solid circles]{Ric06},
\citet[][2SLAQ+SDSS; open diamonds]{Cro09}, \citet[][NDWFS+DLS;
stars]{Gli10}, and \citet[][COSMOS; crosses]{Mas12}.  We have adopted
the following transformation between filters
\citep{Wol03,Ric06,Cro09}:
\begin{eqnarray}
  M_i(z=2) &=& M_g(z=2) - 0.25 \\
  &=& M_{1450} - 0.29 \\
  &=& M_{b_J}  - 0.71
\end{eqnarray}
in order to convert all of the measurements to the $M_i(z=2)$ system
for comparison.

The lifetime, $t_Q$, normalization of the $\Mbh-\Mgal$ relation
($\alpha$ in Equation \ref{eqn:mbh_mgal}) and scatter have been fit to
the data at each redshift.  The grey shaded regions mark the $1\sigma$
range of allowed models.  In most panels the formal errors are so
small that the grey band is buried behind the best-fit relation.  The
constraints on the parameters are so strong because the data at $z<4$
samples luminosities both above and below the knee in the LF and
because the formal errors on the LF are small.

For comparison we also show the luminosity function that results from
assuming a power-law relation between quasar luminosity and halo mass,
as has been assumed in many early works \citep[e.g.][]{EfsRee88,
  Car90, WyiLoe02, WyiLoe03, HaiCioOst04, Mar06, Lid06, Croton09,
  She09, BooSch10}.  This model is characterized by two free
parameters, the duty cycle and the normalization of the (power-law)
relation between quasar luminosity and halo mass\footnote{The
  particular model we consider is $L_Q=\gamma M_h^{1.4}$, where
  $\gamma$ is the free normalization and the index, 1.4, was chosen
  from the power-law model of \citet{Croton09}.}.  The fundamental
difference between our model's predictions and these power-law models
is that we explicitly take into account the efficiency of galaxy
formation as a function of mass and redshift (see Figure
\ref{fig:QAM}).  The two models differ less significantly at higher
redshifts for reasons to be discussed below.

\begin{figure}[!t]
\begin{center}
\resizebox{3.8in}{!}{\includegraphics{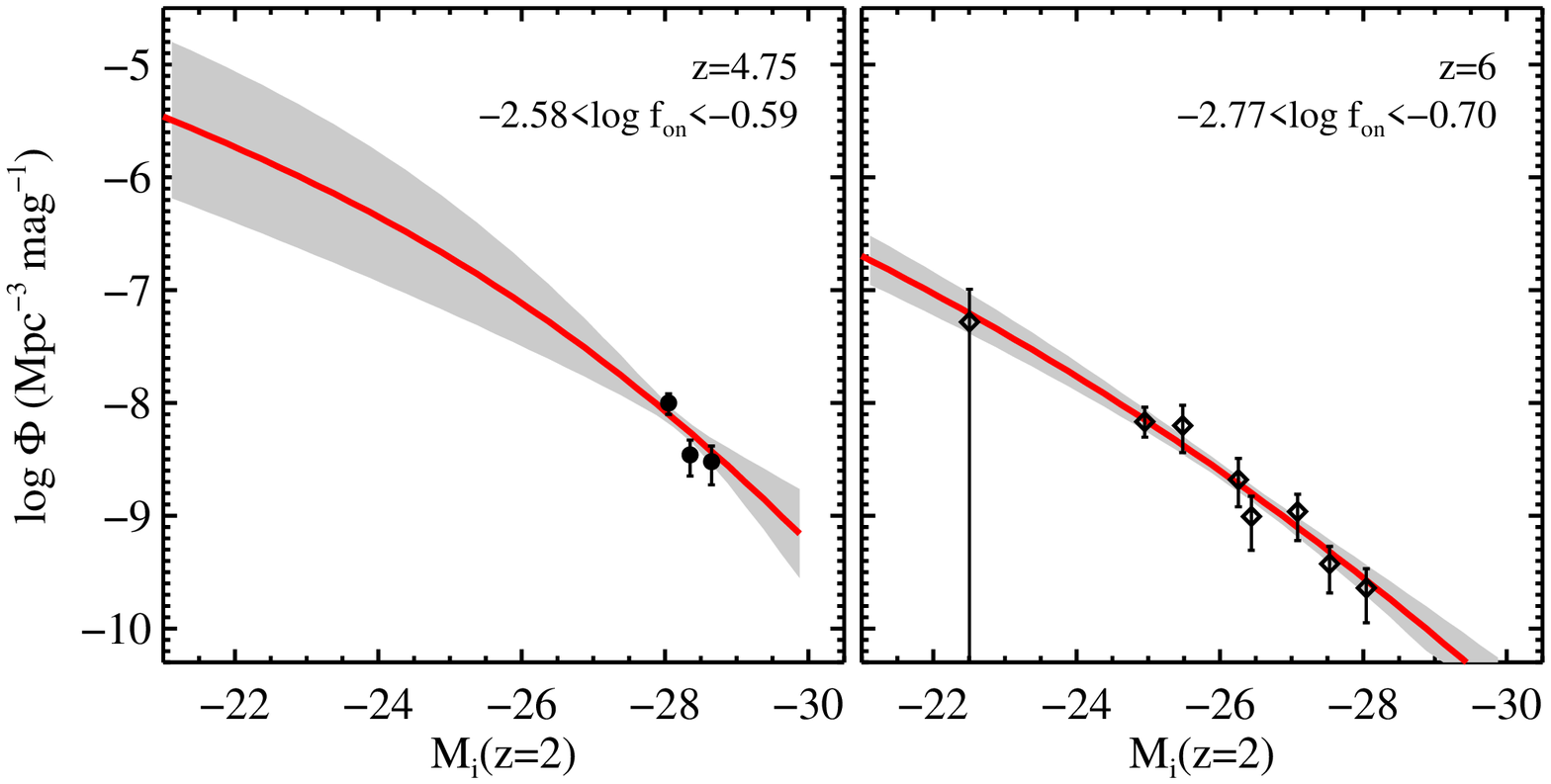}}
\end{center}
\caption{The quasar luminosity function at high redshift.  At $z=4.75$
  the data are from \citet{Ric06} and at $z=6$ the data are from
  \citet{Wil10}.  The best-fit model (solid line) and $1\sigma$
  uncertainty (shaded band) includes variation in the duty cycle,
  normalization in the $\Mbh-\Mgal$ relation and scatter in the
  relation between $\Mgal$ and $L_Q$.  This in contrast to the lower
  redshift fits, where the scatter was held fixed at $0.42$ dex.  At
  high redshift the best-fit scatter exceeds 1 dex.  The $1\sigma$
  range of allowed duty cycles ($f_{\rm on}$) is included in the
  legend in each panel.}
\label{fig:hz}
\end{figure}

In Figure \ref{fig:dutycycle} we show the quasar lifetime, $t_Q$ (or,
equivalently, the duty cycle), the normalization of the $\Mbh-\Mgal$
relation, $\alpha$, and our two model choices for evolution in $\eta$.
In the top panel of Figure \ref{fig:dutycycle} we include lines of
constant duty cycles of $10^{-1}$, $10^{-2}$ and $10^{-3}$.  For
reference, the Salpeter time is the e-folding time for a BH growing at
a fraction $\eta$ of the Eddington luminosity with a radiative
efficiency of $\epsilon$ and is defined as $t_{\rm Salp} = 4\times10^8
(\epsilon/\eta)\,$yr.  It is striking how little $t_Q$ varies from
$0.5<z<3$.  The evidence for a decrease in $t_Q$ at $z>3$ should be
regarded as tentative, as the data used to constrain these parameters
becomes rather uncertain, is compiled from heterogeneous sources, and,
at $z=4.25$, probes a very limited dynamic range.  Moreover, at all
redshifts the formal errors are almost certainly underestimates
because the errors on the observed quasar LFs are only the Poisson
uncertainties, which are vanishingly small for many luminosity
bins. Our estimates of $t_Q$ are in good agreement with quasar
lifetimes inferred by other methods, as summarized in \citet{Mar04}.

In the middle panel of Figure \ref{fig:dutycycle} we show the
evolution of the normalization of the $\Mbh-\Mgal$ relation as
inferred from our model, assuming either a constant or evolving
Eddington ratio.  In this panel we also include the normalization
measured at $z\sim0$ \citep{HarRix04}, and estimates of its evolution
in samples of massive galaxies to $z\sim4$.  The two models produce
very different evolution in normalization of the $\Mbh-\Mgal$
relation, as expected from Equation \ref{eq:lboldef2}.  The model with
constant $\eta$ produces marginally better agreement with the data at
$z<2.5$ although given the likely large systematic uncertainties in
the measurements, it is difficult to draw strong conclusions.  In
particular, scatter in the relation between $\Mgal$ and $L_Q$ can
result in significant biases when inferring mean properties in flux
limited samples \citep{Lau07a, Lau07b}.  Among recent models, the
models of \citet{Hop07a} and \citet{Croton06} predict roughly an order
of magnitude increase in $\Mbh$ at $\Mgal\sim 10^{10}$ between $z=0$
and $z=3$.  In contrast, the simulations of \citet{Sij07} and the
semi-analytic model of \citet{Fan12} predict almost no evolution at
the massive end.

Model fits to the highest redshift quasar LFs are shown separately in
Figure \ref{fig:hz}.  In this case we have included the scatter
between $\Mbh$ and $L_Q$ as an additional free parameter.  This was
necessary because the fiducial model, with a scatter of 0.42 dex,
failed to match the high redshift data without extremely small $f_{\rm
  on}$ and $\alpha$\footnote{We have gone back and re-fit the
  lower-redshift data allowing the scatter to be an additional free
  parameter and found a best-fit scatter that agrees to within
  $\approx0.1$ dex of our fiducial value.  Thus, for simplicity, we
  decided to keep the scatter fixed at 0.42 dex at lower redshifts.}.
For $z=4.75$ and $z=6$ the best-fit scatter is 1.2 and $1.4\,$dex,
respectively.  The $1\sigma$ range of plausible duty cycles, $f_{\rm
  on}$, spans $2\,$dex at these redshifts ($-2.6<{\rm log}\,f_{\rm
  on}<-0.6$ at $z=4.75$ and $-2.8<{\rm log}\,f_{\rm on}<-0.7$ at $z=6$).

Even though the model is not well constrained at high redshift, it is
worth considering these data in some detail.  In particular, if we
focus on $z=6$ we see that the duty cycle is still less than unity and
the scatter in $L_Q-\Mgal$ large.  Our model prefers this solution
because the optically observed quasars are extremely rare ($\Phi\sim
10^{-9}\,{\rm Mpc}^{-3}{\rm mag}^{-1}$) and yet the luminosity
function is not falling exponentially.  If quasars inhabited very high
mass halos and the luminosity was tightly correlated with halo mass
then we would expect an exponential decline at the bright-end of the
luminosity function.  Future constraints on the quasar LF at high
redshift would be very valuable for constraining the duty cycle at
these epochs.  Since rapid accretion rates with long duty cycles seems
to be necessary to produce massive BHs within the first Gyr of cosmic
time, this would provide information on the visibility of this growth
in the resframe ultraviolet and optical.

\begin{figure}[!t]
\begin{center}
\resizebox{3.5in}{!}{\includegraphics{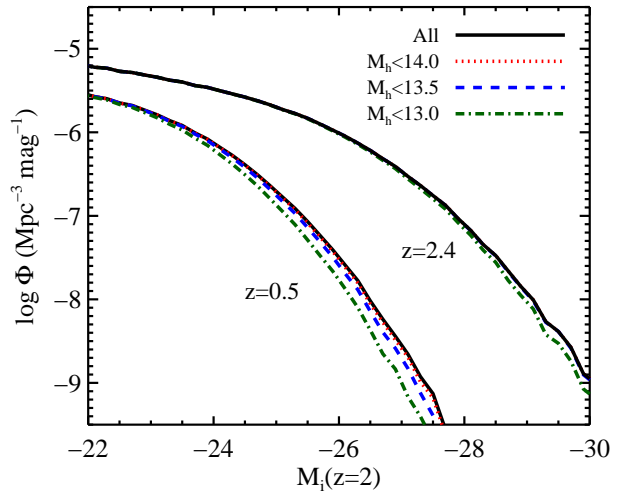}}
\end{center}
\caption{Contribution to the quasar LF from quasars in different halo
  masses.  The curves represent the model LF computed including halos
  less massive than the values shown in the legend (in units of
  log$\Msun$).  The quasar LF is almost entirely insensitive to the
  presence or absence of quasars in halos more massive than
  $10^{13.5}\Msun$.}
\label{fig:qsolfbin}
\end{figure}

Returning to lower redshifts, Figure \ref{fig:qsolfbin} shows the
model LFs at $z=0.5$ and $z=2.4$.  Here we consider the contribution
to the total LF from quasars in halos of different masses.
Specifically, we construct model LFs by selecting quasars residing in
halos less massive than log$(M_h/\Msun)<$ 13.0, 13.5, and 14.0.  The
purpose of this figure is to demonstrate that massive halos contribute
very little to the total LF.  In fact, the model is almost entirely
insensitive to what happens in halos more massive than
log$(M_h/\Msun)<$13.5, owing to their rarity relative to lower mass
halos.  This has important consequences for any model that is tuned to
match the quasar LF, as we discuss in $\S$\ref{sec:discussion}.

\begin{figure}[!t]
\begin{center}
\resizebox{3.7in}{!}{\includegraphics{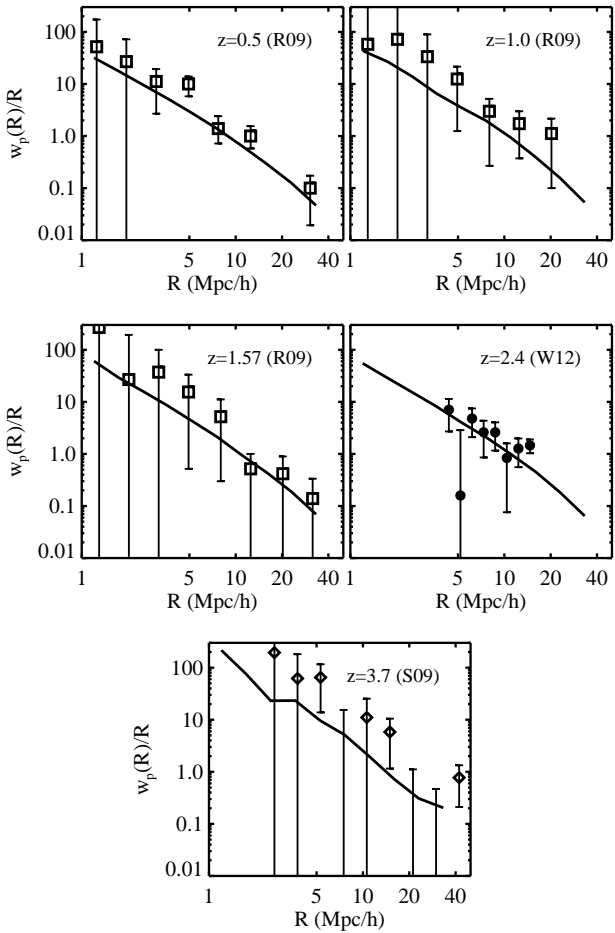}}
\end{center}
\caption{The projected correlation function, $w_p(R)$, vs. projected
  distance, $R$, at 5 redshifts chosen to be representative of the
  data.  We include results from \citet[][R09]{Ros09},
  \citet[][W12]{Whi12}, and \citet[][S09]{Shen++09}, all of which are
  based on data from the Sloan Digital Sky Survey.  At the highest
  redshift there is some tension between the model and data, but the
  error bars are large and the simulation box is too small to provide
  model predictions at the largest scales.  Future measurements of the
  clustering of both low and high redshift quasars will provide
  powerful constraints on the model.}
\label{fig:wp}
\end{figure}

In Figure \ref{fig:lf} we adopted our fiducial values for the slope of
the $\Mbh-\Mgal$ relation.  We found that we can find equally good
fits if we modify the slope of the $\Mbh-\Mgal$ relation to
$\beta=4/3$ or $5/3$, or even if we change the overall normalization
in the $\Mgal-M_h$ relation.  These changes result in different
best-fit values for $t_Q$ and $\alpha$.  Future constraints on the
$\Mbh-\Mgal$ relation as a function of redshift will, in the context
of our model, provide strong constraints on the evolution of the
scatter and the mean Eddington ratio.  Within the parameter space
allowed by the data there are several degeneracies.  For example, an
increase in $t_Q$ can compensate an increase in scatter in the
$L_Q-M_h$ relation.  Increased scatter can also be compensated by
decreasing $\alpha$.  Finally, increasing $\alpha$ can be compensated
by decreasing $t_Q$.

\subsection{Quasar Clustering}
\label{sec:clust}

With the model parameters constrained by the quasar LF, we are now
able to make predictions for the clustering of quasars as a function
of luminosity and redshift.  Recall that our model is characterized by
two parameters, the quasar lifetime, $t_Q$ and the normalization of
the $\Mbh-\Mgal$ relation, $\alpha$.  In the model, we assume that
quasars are a random sample of the BHs in halos, and therefore $t_Q$
has no effect on the clustering of quasars.  The clustering is quite
weakly dependent on the scatter over the luminosity range probed by
current and future planned surveys.  The clustering is therefore only
sensitive to $\alpha$, and this parameter is well-constrained at $z<4$
(see Figure \ref{fig:dutycycle}).  Moreover, $\alpha$ has an
increasingly minor effect on the predicted clustering at higher
redshifts.

Figure \ref{fig:wp} shows a comparison of our model and the data on
the projected autocorrelation function, $w_p(R)$, as a function of
projected (comoving) distance, $R$, for a variety of redshifts chosen
to illustrate the current constraints.  We have computed the model
correlation function by populating the halos drawn from an N-body
simulation\footnote{The simulation employed $2048^3$ particles in a
  cubic box of side length 1 Gpc with a force softening of $14\,$kpc
  (comoving) and was run with the TreePM code of
  \protect\citet{TreePM}.  Halos were found with a friends-of-friends
  algorithm \citep{DEFW} with a linking length of 0.168 times the mean
  inter-particle spacing.  Spherical over-density masses were computed
  for each halo (including a correction for finite resolution).  For
  the range of halo masses and redshifts of interest, masses defined
  via $180\times$ the background density are almost identical to the
  `virial' definition employed by \protect\cite{BehWecCon12}.} with BHs
using the best-fitting relations derived above, and then calculating
the clustering of BHs within the luminosity range of each
observational sample.  This allows us to take into account the
scale-dependent bias and non-linearities, which are important on Mpc
scales.

\begin{figure}[!t]
\begin{center}
\resizebox{3.5in}{!}{\includegraphics{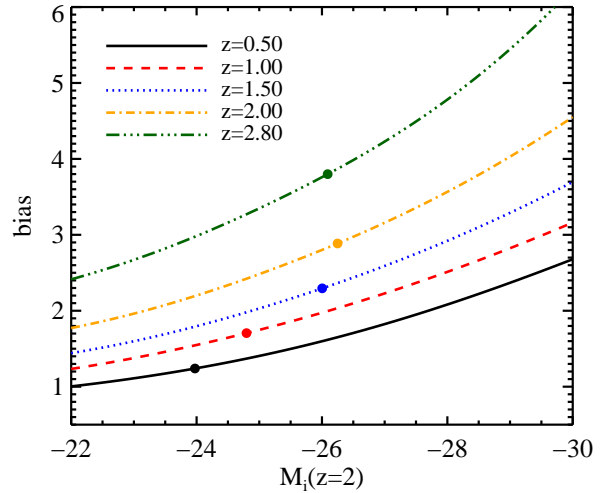}}
\end{center}
\caption{The large-scale bias predicted by our model as a function of
  luminosity for a number of redshifts.  The relation is shallow at
  low luminosity due to the steepness of the $\Mbh-M_h$ relation at
  low mass (see Figure \ref{fig:mbh_mhalo}).  The steepness of the
  relation at high luminosity depends on the scatter in the model,
  being less steep for more scatter.  We have marked on the curves
  where the quasar number density is $5\times10^{-7}{\rm Mpc}^{-3}$,
  which corresponds to of order 100 quasar pairs within $20\,$Mpc in a
  survey volume of $10^{10}\,{\rm Mpc}^{3}$.  To accurately measure
  the bias of objects at lower space densities (and brighter
  luminosities) one would need to resort to cross-correlations.}
\label{fig:bias}
\end{figure}

The majority of models assume that quasar activity occurs due to the
major merger of two gas-rich galaxies, since this scenario provides
the rapid and violent event needed to funnel fuel to the center of the
galaxy \citep[e.g. via the bars-within-bars instability;][]{Shl89} and
feed the central engine while at the same time providing a connection
between BH fueling and the growth of a spheroidal stellar component
\citep[e.g.,][]{Hop08}.  In computing the clustering of quasars we
have populated the halos in the simulation at random, neglecting any
properties of the halos apart from their mass (e.g., whether they have
had a recent major merger).  However, the probability that a halo will
undergo a major merger in a short redshift interval is only weakly
dependent on the mass of the halo \citep{LacCol93, Per03, CohWhi05,
  WetCohWhi09, FakMa09, Hop10b}, i.e., the mass function of such halos
is almost proportional to the mass function of the parent population.
Moreover, the clustering properties of recently merged halos are
similar to a random sample of the population with the same mass
distribution \citep{Per03, WetCohWhi09}.  Thus, our procedure for
randomly selecting halos is consistent with (though not a strong
argument in favor of) the major merger scenario for quasar triggering.

\begin{figure}[!t]
\begin{center}
\resizebox{3.5in}{!}{\includegraphics{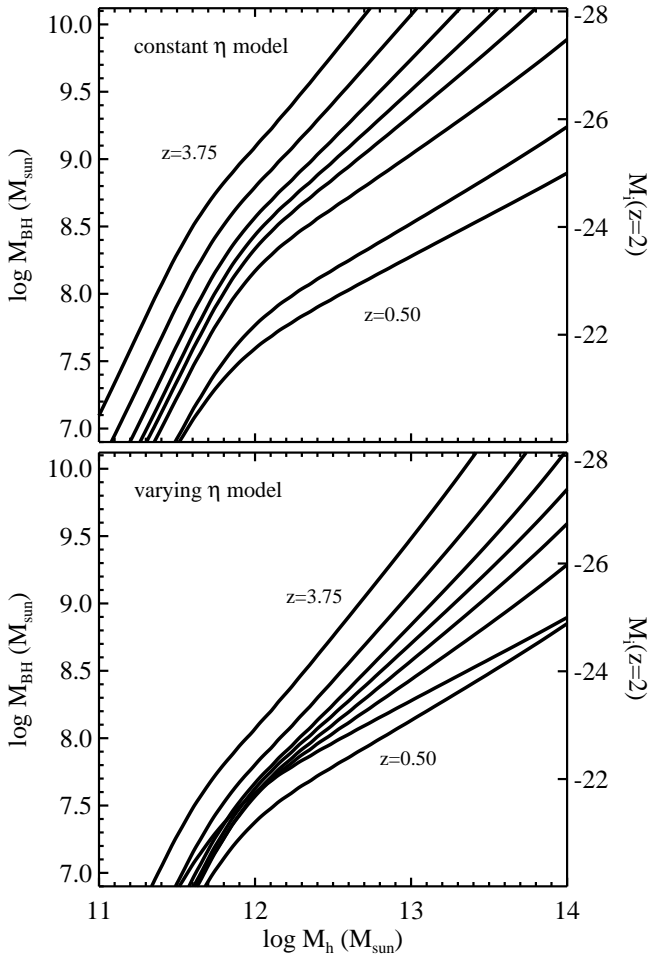}}
\end{center}
\caption{The typical black hole mass in the central galaxy of a halo
  of mass $M_h$, vs. $M_h$, for a number of redshifts (corresponding
  to the redshifts shown in Figure \ref{fig:lf}), for a model with a
  constant Eddington ratio, $\eta$ (top panel), and a model with a
  varying $\eta$ (bottom panel).  The typical BH mass corresponding to
  a fixed $M_h$ increases with $z$, as expected.  Note the significant
  curvature in the relation, which arises due to our assumption that
  galaxy properties regulate the size of black holes and the
  well-known inefficiencies of galaxy formation in high and low halo
  masses.}
\label{fig:mbh_mhalo}
\end{figure}

The agreement between the data and the model is excellent at $z<3$,
especially considering that the model was only tuned to the quasar LF.
The inclusion of satellite quasars would slightly increase the model
prediction in the lowest redshift bin ($z\simeq 0.5$), but any
satellite contribution is quite small for the higher redshifts.  The
model under-predicts the observed clustering at $z\sim3.7$, although
the errors on the data are large.  This model prediction is quite
robust: the $\Mbh-M_h$ relation at high redshift becomes very steep
(see Figure \ref{fig:mbh_mhalo}, discussed below), and so even a
significant change in $\alpha$ or $\eta$ changes the clustering only
modestly.  Similarly, changes in the assumed $L_Q-\Mbh$ scatter within
the range $0.3-0.6\,$dex do not significantly alter the predicted
clustering.  This occurs because a change in scatter induces a change
in $\alpha$ that happens to leave the clustering essentially
unchanged. Future constraints on the clustering of high-redshift
quasars will place strong constraints on this model, as discussed
further in $\S$\ref{sec:imp}, and may indicate that some of our model
assumptions break down as we approach an era of rapid BH growth at
high $z$.

Observationally, it has proven very difficult to measure a dependence
of clustering strength on quasar luminosity \citep[see e.g.,][for a
recent example]{Shen++09}, in part because the significant scatter
between quasar luminosity and halo mass will dilute any intrinsic
relation between clustering strength and luminosity.  We address this
issue in Figure \ref{fig:bias}, where we plot the large-scale bias as
a function of luminosity and redshift.  Here the model bias was
computed via the relation between bias, halo mass, and cosmology from
\citet{Tin10}.

We find a very shallow relation between bias and quasar luminosity
below $M_i(z=2)\sim-26$.  In our model this occurs for three reasons:
(1) the intrinsic relation between bias and halo mass is very shallow
below the characteristic halo mass, which at $z\sim0$ is
$\sim10^{13}\Msun$; (2) the $\Mbh-M_h$ relation becomes very steep at
low mass, implying that a large range in quasar luminosities maps into
a small range in halo masses; (3) scatter in the $\Mgal-M_h$,
$\Mbh-\Mgal$, and $L_Q-\Mbh$ relations dilutes the strong clustering
in high mass halos.  The degree of luminosity dependence (as well as
the absolute value of the bias) is sensitive to the scatter in the
$L_Q-M_h$ relation, with more scatter leading to less $L$-dependence.
This weak luminosity-dependent clustering is also predicted in the
models of \citet{Hop08}, \citet{Croton09} and \citet{She09}.

Figure \ref{fig:bias} demonstrates that we expect significant
luminosity dependent quasar bias only for very luminous quasars.
However, measuring the autocorrelation function of such luminous
quasars is made difficult by their low space densities, which can be
illustrated as follows. The error on the bias in the high-$L$ regime
is dominated by counting statistics.  The number of pairs within e.g.,
$20\,$Mpc is $(1/2)\bar{n}_Q^2\left[1+\bar{\xi}_{20}\right]V_{\rm
  survey}V_{20}$ where $V_{20}=(4\pi/3)(20\,{\rm Mpc})^3$, $V_{\rm
  survey}$ is the survey volume, $\bar{n}_Q$ is the quasar space
density, and $\bar{\xi}$ is the volume average correlation function.
For $\xi(r)=(r_0/r)^2$ we have $\bar{\xi}=3\xi$, and $r_0\sim
10-20\,h^{-1}$Mpc so we expect $\bar{\xi}\sim \mathcal{O}(1)$.  One
hundred pairs within $20\,$Mpc would return an error on the bias of
$\sim10\%$, and for a fiducial survey volume of $10^{10}{\rm Mpc}^3$,
this corresponds to a quasar number density of
$\approx5\times10^{-7}\,{\rm Mpc}^{-3}$.  The luminosity corresponding
to this number density at each redshift is marked by a solid symbol
along the $b(L)$ relation in Figure \ref{fig:bias}.  In order to probe
the bias for quasars at higher luminosities it will be necessary to
resort to cross-correlation techniques, which allow estimates of the
bias of objects with extremely low space density.  An appealing method
would be to cross-correlate existing spectroscopic samples of quasars
with samples of galaxies or lower luminosity quasars selected from
deeper photometry in upcoming surveys such as DES, Pan-STARRS, SUMIRE
and LSST.


\section{Discussion}
\label{sec:discussion}

\subsection{Implications}
\label{sec:imp}

The success of our model in reproducing the basic demographics of
quasars allows us to consider several implications that follow
naturally within our framework.

In Figure \ref{fig:mbh_mhalo} we show the best-fit model $M_h-\Mbh$
relations from $z=0.5$ to $z=3.75$ (the relations above $z=3.75$ are
highly under-constrained and so are not plotted).  As discussed above,
the quasar LF places very weak constraints on the model relations at
$\log(M_h/\Msun)>13.5$, and so one should interpret the model
relations in Figure \ref{fig:mbh_mhalo} with this in mind.  It is also
worth pointing out that while the model formally allows for the
existence of extremely massive BHs with $\Mbh>10^{10}\Msun$ residing
within moderately massive halos, at high redshift such halos are very
rare.  For example, at $z=4.75$ one expects only of order one halo
with log$(M_h/\Msun)>$13 per $10^9$ Mpc$^3$.

\begin{figure}[!t]
\begin{center}
\resizebox{3.5in}{!}{\includegraphics{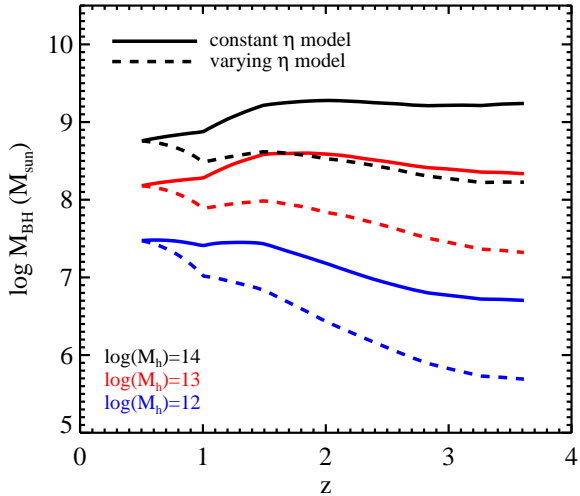}}
\end{center}
\caption{BH growth in the best-fit model from $z=3.75$ to $z=0.5$.
  Results are shown for two choices for the evolution in $\eta$ (see
  the lower panel of Figure \ref{fig:dutycycle}).  Notice that the
  constant $\eta$ model produces massive BHs that lose mass at
  $z<1.5$, suggesting that one or more of the assumptions of this
  model are breaking down at low redshift.  In contrast, the varying
  $\eta$ model produces realistic BH growth at all epochs.  In both
  models lower mass BHs grow more at late times compared to higher
  mass BHs, a phenomenon sometimes referred to as BH downsizing.}
\label{fig:mbh_growth}
\end{figure}

With average mass accretion histories for halos, we can evolve halos
and hence their black holes through the relations shown in Figure
\ref{fig:mbh_mhalo}.  To do this we employ mass accretion histories
presented in \citet{BehWecCon12}, which provide excellent fits to the
results of $N-$body simulations.  The resulting evolution in BH mass
is shown in Figure \ref{fig:mbh_growth} for three representative halo
masses, and for both model choices for the evolution in the Eddington
ratio.  In the model lower mass black holes are growing to lower
redshift faster than higher mass black holes (this is sometimes
referred to as BH downsizing). In the model with a constant $\eta$,
the BHs in the most massive halos lose mass below $z\approx 1.5$,
while in the varying $\eta$ model all BHs grow, if only modestly, at
all epochs.  This suggests that a model with evolving Eddington ratios
may be necessary to ensure self-consistent evolution.  Models that
enforce self-consistent growth of BHs should shed further light on
this problem \citep[e.g.,][]{Mer04, MerHei08, Sha09}.

\begin{figure}[!t]
\begin{center}
\resizebox{3.5in}{!}{\includegraphics{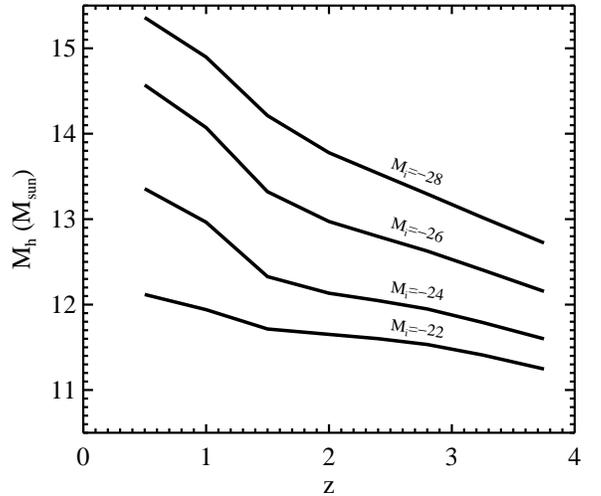}}
\end{center}
\caption{Relation between halo mass and redshift for quasars of a
  fixed luminosity.  At low redshift the range of halo masses hosting
  quasars is very broad, but the distribution narrows substantially at
  high redshift.  This is simply a recasting of the relations shown in
  Figure \ref{fig:mbh_mhalo}.}
\label{fig:mag_mhalo}
\end{figure}

Figure \ref{fig:mag_mhalo} shows the evolution of the halo mass for
quasars of fixed luminosity.  The trend of lower $M_h$ at higher $z$
was already apparent in Figure \ref{fig:mbh_mhalo}.  Figure
\ref{fig:mag_mhalo} also emphasizes how the range of halo masses for a
fixed luminosity range narrows towards higher $z$.  This effect is in
the opposite sense to models which tie the luminosity of quasars
directly to halo properties \citep[e.g.][]{Croton09}.  Our model is
able to reproduce the observed $L$-independent clustering at low $z$
because the run of bias with halo mass also becomes shallower at low
$z$ for the halo masses of interest.

The evolution of the LF shown in Figure \ref{fig:lf} is driven by
evolution in the $\Mbh-\Mgal$ and $\Mgal-M_h$ relations and the
evolution of the halo mass function (evolution in the $L_Q-\Mbh$
relation is governed by evolution in $\eta$).  The break in the model
quasar LF arises primarily due to the shape of the $M_{\rm gal}-M_h$
relation, and thus $L_\star$ quasars live in halos near the peak of
that relation, $M_h\sim 10^{12}M_\odot$.  The peak of the $\Mgal-M_h$
relation changes very little with redshift \citep[e.g.,]{BehWecCon12},
so that at fixed $M_h$ there is little change of $\Mgal$ with $z$.
However the luminosity of the break can evolve due to a combination of
evolution in the $\Mbh-\Mgal$ relation or the Eddington ratio.  In our
fiducial model $\eta$ is constant and $\Mbh\propto (1+z)^2$ at fixed
$M_{\rm gal}$ and so the break in the luminosity function scales as
$(1+z)^2$.  The faint-end slope of the model LF does not vary
significantly, in good agreement with the data, and the overall
normalization changes only modestly.  The major departure from pure
luminosity evolution is the change in the slope of the bright end.
The bright-end slope appears shallower at higher $z$ both because the
data are probing closer to the (brighter) break of the LF and because
the $\Mbh-M_h$ relation becomes steeper at higher mass and redshift.
We also note that the bright end of the model LF is strongly
suppressed at $z<1.5$, and it is this suppression that is responsible
for much of the drop in the quasar number density to lower redshift.
The drop is a consequence of evolving Eddington ratios and the
shallowing of the $\Mbh-M_h$ relation at high mass, which is in turn
driven by the very slow growth of massive galaxies at low redshift.

In fact, the model naturally reproduces the global rise and fall of
the quasar number density over the interval $0.5<z<4.75$.  This
follows simply from the evolution in the $\Mgal-M_h$ and $L_Q-\Mgal$
relations and the halo mass function; it does not require strong
evolution in $t_Q$ at low $z$.  Specifically we do not invoke a
decline in the cold gas fraction nor a decline in the major merger
rate at $z<2$ in order to reproduce the observed decline in the
abundance of quasars.  While these physical processes may ultimately
be responsible for shaping the evolving relations between $L_Q$,
$\Mbh$, $\Mgal$ and $M_h$, they do not appear explicitly in the model.

Our model favors a different picture of how quasars inhabit massive
halos compared to previous work.  Rather than having a preferred halo
mass scale (around $10^{12}\,M_\odot$) for quasar activity, the
present model allows for actively accreting black holes in a broad
range of galaxy and halo masses.  The apparent preference for quasars
to live in halos of $10^{12}\,M_\odot$ arises from the shape of the
$\Mgal-M_h$ relation, which reflects the well known fact that galaxy
formation is most efficient in halos near $10^{12}M_\odot$, along with
the shape of the halo mass function.  Specifically, above the knee in
the $\Mgal-M_h$ relation halos become exponentially rare, while below
the knee a large range in $\Mgal$ maps into a small range in $M_h$.
Thus, the {\it average} halo mass of quasars will be close to the
knee, despite the fact that quasars occupy a broad distribution of
halo masses.

Due to its simplicity the model predicts the clustering of any
population of quasars once the model parameters are fixed (e.g., by
the observed LF).  Variation in the $L_Q-M_h$ scatter or $\Mbh-M_{\rm
  gal}$ slope do not strongly affect the predicted clustering, meaning
that our model makes an essentially parameter-free prediction of the
clustering of quasars as a function of luminosity and redshift.
Overall the agreement between the predicted clustering and the
observations is good, though there is a tendency for the model to
slightly underpredict the observations and there is some tension at
the highest redshifts.  This tension has been noted before -- the very
high amplitude of clustering measured at $z\sim 4$, in combination
with the abundance, requires quasars to have a duty cycle approaching
unity and almost no scatter in $L_Q$ at fixed $M_h$
\citep{WhiMarCoh08, ShaWeiShe10}.  This is at odds with the very low
number densities but power-law decline seen in the luminosity function
at high $z$.  If the clustering measurements can be strengthened,
possibly by cross-correlation of existing spectroscopic quasar samples
with deeper photometric quasar or galaxy samples, then it will
indicate that one of our assumptions is breaking down as we approach
the era of rapid black hole growth in the early Universe.

We make no assumption about what triggers quasar activity, whether it
be a major merger of two gas rich galaxies, a secular instability in a
disk, or a critical halo mass.  In general it is quite difficult to
translate abundance and clustering measurements into constraints on
the underlying mechanisms that trigger quasar activity.  We can gain
some insight by the fact that our duty cycle, or quasar lifetime, is
relatively independent of redshift with a tendency to fall towards
higher redshifts rather than rise.  If quasars are visible for a
fixed, but short, time and are triggered by mergers then we expect
$t_Q$ to scale with the merger rate \citep[c.f.][]{Car90}.  The merger
rate for halos, per halo, per unit redshift is relatively flat
\citep{LacCol93, Per03, CohWhi05, WetCohWhi09, FakMa09, Hop10b}, so if
we can naively translate halo mergers into galaxy mergers we expect a
rate (per unit time) scaling as $(1+z)H(z)\propto (1+z)^{5/2}$ for
$z\gg 1$.  If quasars are visible for a constant interval after each
merger then $t_Q\propto 1+z$, which is not in good agreement with our
best-fit relation.  Of course, galaxy merger rates can differ from
halo merger rates.  A recent analysis by \citet{Hop10a} suggests a
rate per unit time scaling as $(1+z)^{1.5-2.0}$, which would lead to
slower evolution in $t_Q$, as we observe.  Such agreement is not
conclusive however, and we cannot rule out secular processes or a
time-varying combination of multiple triggers.

\subsection{Comparison to Previous Work}

The success of our model in explaining the basic demographics of
quasars with relatively few, smoothly varying inputs goes a long way
to explaining the manner in which forward modeling of the quasar
population can succeed with relatively little fine tuning.  Both
semi-analytic models \citep[e.g.,][]{CatHaeRee99, KauHae00, KauHae02,
  VolHaaMad03, BroSomFab04, Granato04, Croton06, MonFonTaf07, Mal07,
  Bon09, Fan12, Hir12} and hydrodynamic simulations
\citep[e.g.,][]{Sij07, DeGraf11} adjust their subgrid models to ensure
a reasonable match to the $\Mgal-M_h$ relation over a broad redshift
range, thus ensuring that galaxies populate halos in approximately the
correct manner.  All of the models introduce a $\Mbh-\Mgal$ relation
through either or a combination of common feeding mechanisms and
feedback-limited BH growth.  As we have shown, with these two
ingredients even simple lightcurve models are sufficient to match the
basic demographics of quasars over a broad range of luminosity and
redshift.  A good match to the data can be found for a wide range of
scatter in $\Mbh-\Mgal$, or evolution in the scatter.  Conversely, if
a model has difficulties reproducing the stellar mass function and its
evolution then it will need to incorporate mass-dependent quasar
physics that counteracts this deficiency in order to match the
observed quasar properties.

By contrast, models that tie black hole properties directly to the
underlying halo population need to introduce more complexity in order
to reproduce the observed properties of quasars.  Recent examples
include \citet{Lid06}, \citet{Croton09}, and \citet{She09}, who all
need to include mass- and redshift-dependent duty cycles to explain
the shape and evolution of the quasar luminosity function.  While our
model and theirs can produce qualitatively similar fits to the basic
data, the explanations for the observed behaviors differ.  One of the
most basic differences is the range of halos that host active quasars,
and its evolution (discussed above).  This in turn affects how each
model explains the evolution of the quasar LF and the
luminosity-independence of quasar clustering.

Conventional wisdom is that the quasar duty cycle is required by the
data to be a (strong) function of luminosity \citep[e.g.][]{Ade05,
  Hop05, Lid06, Croton09, She09}.  In our model this is not the case.
There are two major reasons for this.  The first is that we obtain a
flattening of the $b(L)$ relation from the steepness of the $L_Q-M_h$
relation at low $L_Q$ and the second is the intrinsic scatter
\footnote{This scatter may arise due to time-dependent processes,
  i.e.~a high $L_Q$ object at the time of observation is not required
  to have always been or continue to be high $L_Q$.}  in that
relation. Thus our model is not a ``light bulb'' model in the sense of
\citet{Hop05} and \citet{Lid06}, who reserve that term for a model in
which there is no scatter in $L_Q-M_h$.  However scatter in the
$L_Q-M_h$ relation is {\it expected\/}, due to the observed scatter in
$\Mbh-\Mgal$ and variation in Eddington ratios if from no other
source; for this reason we refer to our model as a ``scattered'' light
bulb model.  This expected level of scatter is enough to make $b(L)$
flat until extremely high $L$ or correspondingly low $\bar{n}_Q$
\citep[a similar behavior is seen in the model of][which is also not
strictly a light bulb model in the above sense]{Croton09}.  For this
reason we are able to obtain a model in which both the quasar lifetime
and the quasar clustering are independent of $L$.

\citet{Aird12} studied $X-$ray selected active galactic nuclei (AGN)
as a function of galaxy mass at $z\sim0.6$ and found no preference for
AGN to be found in galaxies of a particular mass at fixed Eddington
ratio, even for ratios as high as $\eta\gtrsim0.1$.  Their results
suggest a duty cycle that does not depend strongly on galaxy mass, in
excellent agreement with our results.

Finally, the apparent preference for quasars to live in halos of
$10^{12}M_\odot$, which has been noted by many authors, arises in our
model from the shape of the $\Mgal-M_h$ relation, which reflects the
well-known fact that galaxy formation is most efficient in halos of
$10^{12}M_\odot$, in combination with the halo mass function.  Within
the context of our model this cannot be taken as evidence for a merger
driven origin to quasar activity, despite the fact that it is close to
the small group scale where mergers may be more efficient, because it
is not believed that the knee of the $\Mgal-M_h$ relation is related
to mergers.

\subsection{Mock Catalogs}

While our intent has been to understand the quasar phenomenon, the
model can also be used for the creation of mock catalogs from N-body
simulations.  The simplicity of the model makes it easy to rapidly
generate redshift-dependent quasar populations that have the correct
luminosity function and clustering, given halo catalogs at the
redshifts of interest.  The steps for creating such a catalog are
straightforward:

\begin{itemize}

\item[1.] Adopt the redshift-dependent $\Mgal-M_h$ relation from
  \citet{BehWecCon12}, including scatter in $\Mgal$ at fixed $M_h$.

\item[2.] Use the $\Mgal-\Mbh$ relation from Equation
  \ref{eqn:mbh_mgal} to assign BHs to galaxies, including 0.3 dex of
  scatter in $\Mbh$ at fixed $\Mgal$.  Fix the normalization of this
  relation to the local value, with no redshift evolution (because we
  advocate using the varying $\eta$ model; see below).

\item[3.] Randomly turn a fraction, $f_{\rm on}$, of the BHs into
  active quasars.  As evident from Figure \ref{fig:dutycycle}, the
  quasar lifetime is approximately constant at $3\times10^7$ yr at
  $z<3$; we therefore advocate fixing $t_Q$ to this value.  One then
  determines the duty cycle via $f_{\rm on}(z)=t_Q/t_H(z)$.

\item[4.] For the active BHs, convert $\Mbh$ into $L_Q$ using Equation
  \ref{eqn:Lq_Mbh}, with an additional 0.3 dex of scatter in $L_Q$ at
  fixed $\Mbh$.  Use the redshift-dependent Eddington ratio, $\eta$,
  shown in the bottom panel of Figure \ref{fig:dutycycle}.  We
  advocate using the varying $\eta$ model because this model produces
  self-consistent BH growth at all redshifts (see Figure
  \ref{fig:mbh_mhalo}).

\end{itemize}

When simulations are populated with quasars in this way, the mock
quasar LF and clustering will agree with all existing LF and
clustering data at $z<3$.  In order to produce mock catalogs at higher
redshifts one will need to include a drop in $t_Q$ as shown in Figure
\ref{fig:dutycycle}.  Such mock catalogs should prove useful in the
context of ongoing and future planned surveys such as BOSS, bigBOSS,
DES, Pan-STARRS, SUMIRE and LSST.


\section{Summary}
\label{sec:conclusions}

We have presented a simple model for quasars with the aim of
understanding to what extent their demographics arise naturally from
what is known about the evolution of galaxies, along with plausible
assumptions about how black holes inhabit them.  The key feature of
the model is that the properties of black holes are set by those of
their host galaxies rather than their host halos \citep[see
also][]{Whi12}.  In the model, BH mass is linearly related to galaxy
mass and BHs shine at a fixed fraction of the Eddington luminosity
during accretion episodes.  Galaxies are related to dark matter halos
via empirically constrained relations \citep{BehWecCon12}.  The model
has only two free parameters at each redshift, the normalization of
the $\Mbh-\Mgal$ relation and the duty cycle, both of which are
tightly constrained by observations of the quasar LF.  We have
explored two possibilities for the evolution of the Eddington ratio
with redshift, finding physically self-consistent BH growth for a
model in which the Eddington ratio increases with increasing redshift.
The model provides an excellent fit to the LF data for $0.5<z<6$ and
reproduces the observed clustering at intermediate redshifts with no
additional adjustable parameters.

The best-fit model parameters imply a quasar lifetime of approximately
$3\times10^7\,$yr at $z<3$.  This may be expected if the growth of the
galaxy during a quasar event only allows $\sim 1$ e-folding of black
hole growth before feedback halts quasar activity.

There are several implications of our model, which we now summarize:

\begin{itemize}

\item Actively accreting BHs are equally likely to exist in galaxies,
  and dark matter halos, over a wide range in masses.  The BHs in
  halos more massive than $10^{13.5}M_\odot$ contribute very little to
  the observed quasar LF at any redshift due to their rarity.  The
  quasar LF therefore places weak constraints on the quasar duty cycle
  in massive halos.  

\item The break in the quasar LF is a reflection of the break in the
  $\Mgal-M_h$ relation at $M_h\sim10^{12}M_\odot$ and the observed
  evolution of the LF primarily reflects the $(1+z)^2$ scaling of
  $L_Q/\Mgal$ and the change in shape of the $\Mgal-M_h$ relation.
  The bright-end slope of the quasar LF appears shallower at high $z$
  both because the data are probing closer to the (brighter) break in
  the LF and because the $\Mbh-M_h$ relation becomes steeper at higher
  mass and redshift.

\item Our model naturally reproduces the global rise and fall of the
  quasar number density over the interval $0.5<z<6$.  This follows
  simply from the evolution in the $L_Q-M_h$ relation and does not
  require strong evolution in the quasar lifetime at $z<3$.  The
  bright end of the model quasar LF is strongly suppressed at $z<1.5$,
  due to the slow growth of massive galaxies, and this is responsible
  for much of the drop in quasar number density to low redshift.

\item The apparent preference for quasars to live in halos of
  $10^{12}M_\odot$ arises from the shape of the $\Mgal-M_h$ relation,
  which reflects the well-known fact that galaxy formation is most
  efficient near $10^{12}M_\odot$, in conjunction with the steepness
  of the halo mass function at high mass.

\item There is some tension between our model and the amplitude of
  clustering observed at $z\sim 4$; the latter, taken at face value,
  suggests that quasars have a duty cycle approaching unity and almost
  no scatter in $L_Q-M_h$ while the power-law fall-off of the bright
  end of the luminosity function suggests otherwise.  Future
  clustering measurements in this redshift range will be crucial tests
  of the model.

\item The nearly constant inferred quasar lifetimes as a function of
  luminosity and redshift (at $z<3$) should provide valuable
  constraints on the triggering mechanisms for quasars.

\end{itemize}

Measurements of quasar demographics at higher redshifts and lower
luminosities will help to further constrain and test our model.  In
particular, stronger constraints on the quasar LF at $z>4$, on quasar
clustering as a function of luminosity and redshift, and on the
$\Mbh-\Mgal$ relation as a function of redshift, will provide very
strong constraints on the model parameters.  Moreover, with such
observational constraints in hand, we will be able to directly
constrain the mean Eddington ratio as a function of redshift and the
scatter as a function of redshift, providing further insight into the
link between quasars, black holes, galaxies, and dark matter halos.


\acknowledgments 

We thank Nic Ross and Yue Shen for providing their data in electronic
form, Adam Myers, Matt McQuinn, and Yue Shen for comments on an
earlier draft, and Tom Targett for his literature compilation of data
that went into Figure 4.  The referee is thanked for comments that
improved the quality of the manuscript.  M.W. was supported by the NSF
and NASA.  This work made extensive use of the NASA Astrophysics Data
System and of the {\tt astro-ph} preprint archive at {\tt arXiv.org}.
The analysis made use of the computing resources of the National
Energy Research Scientific Computing Center.

\end{document}